\documentstyle[epsfig,12pt,a4p]{article}

\newcommand{\pipipi}{\mbox{$\pi^+\pi^-\pi^0$ }}

\newcommand{\kpikpi}{\mbox{$K^+K^-\pi^+\pi^-$ }}

\newcommand{\pipipipi}{\mbox{$\pi^+\pi^-\pi^+\pi^-$ }}
\newcommand{\pipi}{\mbox{$\pi^+\pi^-$ }}
\newcommand{\ppbar}{\mbox{$p \overline p$ }}
\newcommand{\ppbarpi}{\mbox{$p \overline p \pi^0$ }}
\newcommand{\ppbarpipi}{\mbox{$p \overline p \pi^+\pi^-$ }}
\newcommand{\lamlam}{\mbox{$\Lambda  \overline \Lambda $ }}
\begin{document}
\begin{titlepage}
\def\footnoterule{\hrule width 1.0\columnwidth}
\begin{tabbing}
put this on the right hand corner using tabbing so it looks
 and neat and in \= \kill
\> {3 December 1998}
\end{tabbing}
\bigskip
\bigskip
\begin{center}{\Large  {\bf A study of the centrally produced
baryon-antibaryon systems
in pp interactions at 450 GeV/c}
}\end{center}
\bigskip
\bigskip
\begin{center}{        The WA102 Collaboration
}\end{center}\bigskip
\begin{center}{
D.\thinspace Barberis$^{  4}$,
W.\thinspace Beusch$^{   4}$,
F.G.\thinspace Binon$^{   6}$,
A.M.\thinspace Blick$^{   5}$,
F.E.\thinspace Close$^{  3,4}$,
K.M.\thinspace Danielsen$^{ 10}$,
A.V.\thinspace Dolgopolov$^{  5}$,
S.V.\thinspace Donskov$^{  5}$,
B.C.\thinspace Earl$^{  3}$,
D.\thinspace Evans$^{  3}$,
B.R.\thinspace French$^{  4}$,
T.\thinspace Hino$^{ 11}$,
S.\thinspace Inaba$^{   8}$,
A.V.\thinspace Inyakin$^{  5}$,
T.\thinspace Ishida$^{   8}$,
A.\thinspace Jacholkowski$^{   4}$,
T.\thinspace Jacobsen$^{  10}$,
G.T\thinspace Jones$^{  3}$,
G.V.\thinspace Khaustov$^{  5}$,
T.\thinspace Kinashi$^{  12}$,
J.B.\thinspace Kinson$^{   3}$,
A.\thinspace Kirk$^{   3}$,
W.\thinspace Klempt$^{  4}$,
V.\thinspace Kolosov$^{  5}$,
A.A.\thinspace Kondashov$^{  5}$,
A.A.\thinspace Lednev$^{  5}$,
V.\thinspace Lenti$^{  4}$,
S.\thinspace Maljukov$^{   7}$,
P.\thinspace Martinengo$^{   4}$,
I.\thinspace Minashvili$^{   7}$,
T.\thinspace Nakagawa$^{  11}$,
K.L.\thinspace Norman$^{   3}$,
J.P.\thinspace Peigneux$^{  1}$,
S.A.\thinspace Polovnikov$^{  5}$,
V.A.\thinspace Polyakov$^{  5}$,
V.\thinspace Romanovsky$^{   7}$,
H.\thinspace Rotscheidt$^{   4}$,
V.\thinspace Rumyantsev$^{   7}$,
N.\thinspace Russakovich$^{   7}$,
V.D.\thinspace Samoylenko$^{  5}$,
A.\thinspace Semenov$^{   7}$,
M.\thinspace Sen\'{e}$^{   4}$,
R.\thinspace Sen\'{e}$^{   4}$,
P.M.\thinspace Shagin$^{  5}$,
H.\thinspace Shimizu$^{ 12}$,
A.V.\thinspace Singovsky$^{ 1,5}$,
A.\thinspace Sobol$^{   5}$,
A.\thinspace Solovjev$^{   7}$,
M.\thinspace Stassinaki$^{   2}$,
J.P.\thinspace Stroot$^{  6}$,
V.P.\thinspace Sugonyaev$^{  5}$,
K.\thinspace Takamatsu$^{ 9}$,
G.\thinspace Tchlatchidze$^{   7}$,
T.\thinspace Tsuru$^{   8}$,
M.\thinspace Venables$^{  3}$,
O.\thinspace Villalobos Baillie$^{   3}$,
M.F.\thinspace Votruba$^{   3}$,
Y.\thinspace Yasu$^{   8}$.
}\end{center}

\begin{center}{\bf {{\bf Abstract}}}\end{center}

{
A study of the centrally produced \ppbar, \ppbarpi, \ppbarpipi
and \lamlam channels has been performed in $pp$ collisions using
an incident beam momentum of 450~GeV/c.
No significant new structures are observed in the mass spectra, however,
important new information on the production dynamics is obtained.
A systematic study of the production properties of these systems has
been performed
and
it is found that these systems are not
produced dominantly by double Pomeron exchange.
}
\bigskip
\bigskip
\bigskip
\bigskip\begin{center}{{Submitted to Physics Letters}}
\end{center}
\bigskip
\bigskip
\begin{tabbing}
aba \=   \kill
$^1$ \> \small
LAPP-IN2P3, Annecy, France. \\
$^2$ \> \small
Athens University, Physics Department, Athens, Greece. \\
$^3$ \> \small
School of Physics and Astronomy, University of Birmingham, Birmingham, U.K. \\
$^4$ \> \small
CERN - European Organization for Nuclear Research, Geneva, Switzerland. \\
$^5$ \> \small
IHEP, Protvino, Russia. \\
$^6$ \> \small
IISN, Belgium. \\
$^7$ \> \small
JINR, Dubna, Russia. \\
$^8$ \> \small
High Energy Accelerator Research Organization (KEK), Tsukuba, Ibaraki 305,
Japan. \\
$^{9}$ \> \small
Faculty of Engineering, Miyazaki University, Miyazaki, Japan. \\
$^{10}$ \> \small
Oslo University, Oslo, Norway. \\
$^{11}$ \> \small
Faculty of Science, Tohoku University, Aoba-ku, Sendai 980, Japan. \\
$^{12}$ \> \small
Faculty of Science, Yamagata University, Yamagata 990, Japan. \\
\end{tabbing}
\end{titlepage}
\setcounter{page}{2}
\bigskip
\par
\par
Experiment WA102 is designed to study exclusive final states formed in
the reaction
\begin{equation}
pp \rightarrow p_{f} (X^0) p_{s}
\end{equation}
at 450 GeV/c.
The subscripts $f$ and $s$ indicate the
fastest and slowest particles in the laboratory respectively
and $X^0$ represents the central
system that is presumed to be produced by double exchange processes.
The experiment
has been performed using the CERN Omega Spectrometer,
the layout of which is
described in ref.~\cite{WADPT}.
In previous analyses of other channels it has been observed that
when the centrally produced system has been analysed
as a function of the parameter $dP_T$, which is the difference
in the transverse momentum vectors of the two exchange
particles~\cite{WADPT,closeak},
all the undisputed $q \overline q$ states are suppressed at small
$dP_T$, in contrast to glueball candidates.
\par
This paper presents a study of the
\ppbar,
\ppbarpi,
\ppbarpipi and \lamlam
final states
at a centre of mass energy of
$\sqrt s$~=~29.1~GeV.
The
\ppbar,
\ppbarpipi and \lamlam
channels have previously been studied at
$\sqrt s$~=~12.7~GeV~\cite{oldppbar}.
In recent years there have been claims of the observation
of two different resonant signals in the \ppbar channel.
The first
claim is for the observation of the $\xi(2220)$,
with a width of 20~MeV,
in radiative $J/\psi$ decays made by the BES
collaboration~\cite{BES}. The $\xi(2220)$ is claimed
to be a good candidate for the tensor glueball.
To date,
every established $J^{PC}$~=~$0^{++}$ and $2^{++}$
glueball candidate has been observed in
central $pp$ collisions. Therefore,
it is important to look for these new states in central production
in order to learn more about the nature and/or existence of the $\xi(2220)$.
\par
The second claimed observation is for a state with a mass of
2.02~GeV and a width of less than 10~MeV observed in
central baryon exchange by the WA56 experiment~\cite{VAS}.
It is claimed that this state could be interpreted as
a baryonium candidate. Although the current experiment
does not study baryon exchange it does study central production
and hence a search for this state may be of interest.
\par
In addition to these searches, an analysis of the production kinematics
of baryon-antibaryon production is presented which
can give information on the mechanism of the formation
of these final states in central production.
\par
The reaction
\begin{equation}
pp \rightarrow p_{f} (p \overline p ) p_{s}
\label{eq:b}
\end{equation}
has been isolated
from the sample of events having four
outgoing
charged tracks,
by first imposing the following cuts on the components of the
missing momentum:
$|$missing~$P_{x}| <  14.0$ GeV/c,
$|$missing~$P_{y}| <  0.16$ GeV/c and
$|$missing~$P_{z}| <  0.08$ GeV/c,
where the x axis is along the beam
direction.
A correlation between
pulse-height and momentum
obtained from a system of
scintillation counters was used to ensure that the slow
particle was a proton.
\par
In order to select the \ppbar system, information
from the {\v C}erenkov counter was used.
One centrally produced charged
particle was required to be identified as a $p$ or an ambiguous K/$p$
by the {\v C}erenkov counter
and the other particle was required to be consistent with being a proton.
The method of Ehrlich et al.~\cite{EHRLICH},
has been used to compute the mass
squared of the two centrally produced particles assuming them to have
equal mass.
The resulting distribution is shown in
fig.~\ref{fi:1}a) where a clear peak can be seen at
the proton mass squared. This distribution has been fitted with
Gaussians to represent the contributions from the
\pipi, $K^+K^-$  and
\ppbar channels.
{}From this fit the number of \ppbar events is found to
be 6256~$\pm$~220.
\par
A cut on the Ehrlich mass squared of
$0.65 \leq M^2_X \leq 1.15$~$GeV^2$
has been used to select a sample of \ppbar events.
The resulting \ppbar effective mass distribution is shown
in fig.~\ref{fi:1}b).
There are no significant structures in the mass spectrum, in particular
there is no evidence for the $\xi(2220)$ that has been claimed in the \ppbar
channel by the BES collaboration~\cite{BES} nor is there evidence
for the narrow structure at 2.02~GeV claimed to have been observed
in central baryon exchange reactions~\cite{VAS}.
The mass resolution of the WA102 experiment
in each region is better than~10~MeV.
We can calculate an upper limit for the cross-sections for the production
of these claimed resonances in central $pp$ interactions to be
$\sigma(\xi(2220)) < 1.6$ nb and
$\sigma(2.02 GeV) < 1.4$ nb at 95~\% confidence level.
\par
The reaction
\begin{equation}
pp \rightarrow p_{f} (p \overline p \pi^0) p_{s}
\label{eq:c}
\end{equation}
where the $\pi^0$ has been observed decaying to $\gamma \gamma$,
has been isolated
from the sample of events having four
outgoing
charged tracks
plus two $\gamma$s each with energy greater than 0.5 GeV
reconstructed in the electromagnetic
calorimeter\footnote{The showers associated with the impact of
the charged tracks on the calorimeter
have been removed from the event before the
requirement of only two $\gamma$s was made.},
by first imposing the following cuts on the components of the
missing momentum:
$|$missing~$P_{x}| <  17.0$ GeV/c,
$|$missing~$P_{y}| <  0.16$ GeV/c and
$|$missing~$P_{z}| <  0.12$ GeV/c.
\par
One centrally produced charged
particle was required to be identified as a $p$ or an ambiguous K/$p$
by the {\v C}erenkov counter
and the other particle was required to be consistent with being a proton.
The Ehrlich mass distribution is shown in
fig.~\ref{fi:1}c) where a clear peak can be seen at
the proton mass squared. This distribution has been fitted with
Gaussians to represent the contributions from the
\pipipi, $K^+K^-\pi^0$  and
\ppbarpi channels.
{}From this fit the number of \ppbarpi events is found to
be 877$~\pm$~85.
A cut on the Ehrlich mass squared of
$0.65 \leq M^2_X \leq 1.15$~$GeV^2$
has been used to select a sample of \ppbarpi events.
The resulting \ppbarpi effective mass distribution is shown in
fig.~\ref{fi:1}d) where no significant structure can be observed.
Similarly,
the $p \overline p$, $p \pi^0$ and $\overline p \pi^0$
mass spectra do not show any significant structures (not shown).
\par
The reaction
\begin{equation}
pp \rightarrow p_{f} (p \overline p \pi^+\pi^-) p_{s}
\label{eq:d}
\end{equation}
has been isolated
from the sample of events having six
outgoing
charged tracks,
by first imposing the following cuts on the components of the
missing momentum:
$|$missing~$P_{x}| <  14.0$ GeV/c,
$|$missing~$P_{y}| <  0.12$ GeV/c and
$|$missing~$P_{z}| <  0.08$ GeV/c.
\par
In order to select the \ppbarpipi system
an event was accepted if a positive or negative particle
was identified as a $p$ or a K/$p$ by the {\v C}erenkov system
and the other particle with the
same charge was consistent with being a $\pi$.
A modified method of Ehrlich et al.~\cite{EHRLICH},
has been used to compute the mass
squared of the two highest momentum central particles assuming the other
two particles to be pions.
The resulting distribution is shown in
fig.~\ref{fi:2}a) where a clear peak can be seen at
the proton mass squared. This distribution has been fitted with
Gaussians to represent the contributions from the
\pipipipi,
\kpikpi and \ppbarpipi channels.
{}From this fit the number of \ppbarpipi events is found to
be 2076~$\pm$~160.
A cut on the Ehrlich mass squared of
$0.65 \leq M^2_X \leq 1.15$~$GeV^2$
has been used to select a sample of \ppbarpipi events.
The resulting \ppbarpipi
effective mass spectrum is shown in fig.~\ref{fi:2}b) and
shows a broad distribution with a maximum near threshold.
\par
A study has been performed of the various two and three body
subsystems but no structure has been observed
except $\Delta^{++}$ and $\overline \Delta^{--}$
in the $p \pi^+$ and $\overline p \pi^-$ mass spectra respectively.
Fig.~\ref{fi:2}c) shows a scatter plot of
M($\overline p \pi^-)$
versus M($p\pi^+)$ where a
clear accumulation of events can be observed in the
$\Delta^{++} \overline \Delta^{--}$ region. However,
it is not possible to extract a reliable measure of the
$\Delta^{++} \overline \Delta^{--}$ contribution
due to the difficulties in establishing the level of background.
\par
The reaction
\begin{equation}
pp \rightarrow p_{f} (\Lambda \overline \Lambda) p_{s}
\label{eq:e}
\end{equation}
has been isolated
from the sample of events having two
outgoing
charged tracks plus two $V^0$s,
by first imposing the following cuts on the components of the
missing momentum:
$|$missing~$P_{x}| <  14.0$ GeV/c,
$|$missing~$P_{y}| <  0.16$ GeV/c and
$|$missing~$P_{z}| <  0.02$ GeV/c.
For each $V^0$ the value of
$\alpha=\frac{P_L^+ - P_L^-}{P_L^++P_L^-}$,
was calculated,
where $P_L^+$ ($P_L^-$) is the longitudinal momentum of
the positive (negative) particle from the decay of the $V^0$
with respect to the $V^0$ momentum vector.
For a $\Lambda$ ($\overline \Lambda$) $\alpha$
is positive (negative) and hence events which were compatible
with being \lamlam were selected by requiring that
the product $\alpha_1 . \alpha_2$ was negative.
\par
The quantity $\Delta$, defined as
$ \Delta = MM^{2}(p_{f}p_{s}) - M^{2}(\Lambda \overline \Lambda)$,
where $MM^{2}(p_{f}p_{s})$ is the missing mass squared of the two
outgoing protons,
was then calculated for each event and
a cut of $|\Delta|$ $\leq$ 2.0 (GeV)$^{2}$ was used to select the
\lamlam channel.
In order to study any possible residual $K^0_S$ contamination
a scatter plot of M$(\pi\pi)$ versus M$(p\pi)$ is shown
in fig.~\ref{fi:3}a) for the case when the
other $V^0$ is compatible with being a $\Lambda$
(1.09~$\leq$~M($p\pi$)~$\leq$~1.14~GeV).
The resulting $p\pi^-$~($\overline p \pi^+$) mass
distribution is shown in fig~\ref{fi:3}b) where
a clear $\Lambda$~($\overline \Lambda$) signal can be seen
over little background,
with negligible contribution from the $K^0_S$.
The resulting \lamlam effective mass spectrum is shown in fig.~\ref{fi:3}c)
and consists of 123 events.
\par
A study of the \ppbar, \ppbarpi, \ppbarpipi and
\lamlam systems has been performed as a function of the parameter
$dP_T$, which is the difference in the transverse momentum vectors
of the two exchanged
particles~\cite{WADPT,closeak}.
After acceptance corrections
the results are shown in table~\ref{ta:a} together with
the value of the ratio (R) of events at small $dP_T$ to large $dP_T$.
In previous studies~\cite{memoriam} of the ratio R
we have observed that all systems fall into three distinct classes.
Firstly, there are all the undisputed $q \overline q$ states
which can be produced in Double Pomeron Exchange (DPE),
namely those with positive G parity and $I=0$,
which
have a small value for this ratio ($< 0.1$).
Secondly, there are those
states with $I=1$ or G parity negative,
which cannot be produced by DPE,
which have a slightly higher value ($\approx 0.25$).
Finally, there are the
states which could have a gluonic component, which have
a large value for this ratio ($> 0.6$).
It is interesting to note that the baryon-antibaryon systems
have a value of R consistent with the second class, i.e.
that they are not produced by DPE.
This fact can be investigated by studying the cross-section
dependence as a function of centre of mass energy.
\par
After correcting for
geometrical acceptances, detector efficiencies and
losses due to selection cuts,
the cross-sections for the channels
at $\sqrt s$~=~29.1~GeV in the $x_F$ interval
$|x_F| \leq 0.2$ have been calculated and are shown
in table~\ref{ta:b}.
These can be compared, where possible,
to the cross-sections found at $\sqrt s$~=~12.7~GeV which are also
shown in table~\ref{ta:b}.
As can be seen the cross-sections
are decreasing with increasing centre of mass
energy. This is not consistent with them being produced dominantly
by DPE and suggest that these systems are produced by
double Regge or Regge-Pomeron exchanges~\cite{dpet}.
\par
The acceptance corrected azimuthal angle ($\phi$) between the $p_T$
vectors of the two protons ($p_f$ and $p_s$)
is shown in
fig.~\ref{fi:4}a), c), e) and g).
The distributions in all cases
are consistent with being flat.
Although naively a flat distribution would be expected,
this is the first time that a system or resonance has been
observed to have a flat $\phi$ distribution~\cite{papang}.
\par
Fig.~\ref{fi:4}b), d), f) and h)  shows the four momentum transfer squared at
one of the proton vertices.
The distributions have been fitted with a single exponential
of the form $exp(-b |t|)$ and the results
are presented in table~\ref{ta:b}.
The first bin in the distributions has been excluded from the fit
due to the fact that the uncertainties in the acceptance correction
are greatest in this bin.
\par
In conclusion, a study of the centrally produced
\ppbar, \ppbarpi, \ppbarpipi and \lamlam channels has been performed.
There is no evidence for resonance production with the exception
of $\Delta^{++}$ and $\overline \Delta^{--}$ in the \ppbarpipi channel.
In the \ppbar channel
there is no evidence for the $\xi(2220)$ and
an upper limit on the cross-section
for its production in central $pp$ collisions
has been calculated to be 1.6~nb.
{}A study of the centre of mass energy dependence for the
production of central baryon-antibaryon systems shows that they
are not produced dominantly by double Pomeron exchange.
\begin{center}
{\bf Acknowledgements}
\end{center}
\par
This work is supported, in part, by grants from
the British Particle Physics and Astronomy Research Council,
the British Royal Society,
the Ministry of Education, Science, Sports and Culture of Japan
(grants no. 04044159 and 07044098), the Programme International
de Cooperation Scientifique (grant no. 576)
and
the Russian Foundation for Basic Research
(grants 96-15-96633 and 98-02-22032).

\newpage
\newpage
\begin{table}[h]
\caption{Production of the channels as a function of $dP_T$
expressed as a percentage of its total contribution and the
ratio (R) of events produced at $dP_T$~$\leq$~0.2~GeV to the events
produced at $dP_T$~$\geq$~0.5~GeV.}
\label{ta:a}
\vspace{1in}
\begin{center}
\begin{tabular}{|c|c|c|c|c|} \hline
 & & & & \\
 &$dP_T$$\leq$0.2 GeV & 0.2$\leq$$dP_T$$\leq$0.5 GeV &$dP_T$$\geq$0.5 GeV &
$R=\frac{dP_T \leq 0.2 GeV}{dP_T\geq 0.5 GeV}$\\
 & & & & \\ \hline
 & & & & \\
$p \overline p$  &14.4$\pm$ 0.4 & 44.6 $\pm$ 0.7   &41.0 $\pm$ 0.7 &
0.35~$\pm$~0.01 \\
 & & & & \\ \hline
 & & & & \\
$p \overline p \pi^0$  &14.5 $\pm$ 1.0 & 43.3 $\pm$ 1.6  &42.2 $\pm$ 1.6  &
0.34~$\pm$~0.03\\
 & & & & \\ \hline
 & & & & \\
$p \overline p \pi^+\pi^-$  &13.9 $\pm$ 0.5  & 43.3 $\pm$ 0.9  &42.8 $\pm$ 0.9
 & 0.32~$\pm$~0.01 \\
 & & & & \\ \hline
 & & & & \\
$\Lambda \overline \Lambda$  &15.0 $\pm$ 3.5  & 39.4 $\pm$ 5.6 &45.6 $\pm$ 6.1
& 0.33~$\pm$~0.09\\
 & & & & \\ \hline
\end{tabular}
\end{center}
\end{table}
\newpage
\begin{table}[h]
\caption{The cross-sections and slope parameter of the
four momentum transfer squared
for the \ppbar, \ppbarpi, \ppbarpipi and \lamlam channels.}
\label{ta:b}
\vspace{1in}
\begin{center}
\begin{tabular}{|c|cc|c|} \hline
& & & \\
Channel & \multicolumn{2}{c|}{Cross-section$\setminus$nb} & Slope \\
        & \multicolumn{2}{c|}{ } & b                      \\
        & $\sqrt s$=12.7 GeV & $\sqrt s$=29.1 GeV & GeV$^{-2}$ \\
& & & \\ \hline
& & & \\
$p \overline p$ &400~$\pm$~60 &186~$\pm$~19 &5.5~$\pm$~0.3 \\
& & & \\
$p \overline p \pi^0$ &- &43~$\pm$~5  &5.9~$\pm$~0.5 \\
& & & \\
$p \overline p \pi^+\pi^-$&226~$\pm$~42 &82~$\pm$~7  &5.4~$\pm$~0.3 \\
& & & \\
$\Lambda \overline \Lambda$ &29~$\pm$~8 &12~$\pm$~2  &4.5~$\pm$~2.0 \\
 & & & \\ \hline
\end{tabular}
\end{center}
\end{table}
\clearpage
{ \large \bf Figures \rm}
\begin{figure}[h]
\caption{a) The Ehrlich mass squared distribution and
b) the \ppbar mass spectrum for the \ppbar channel.
c) The Ehrlich mass squared distribution and
d) the \ppbarpi mass spectrum for the \ppbarpi channel.
}
\label{fi:1}
\end{figure}
\begin{figure}[h]
\caption{
For the \ppbarpipi channel
a) the Ehrlich mass squared distribution,
b) the \ppbarpipi mass spectrum and
c) the scatter plot of M($\overline p \pi^-$) versus
M($p \pi^+$).
}
\label{fi:2}
\end{figure}
\begin{figure}[h]
\caption{For the \lamlam channel
a) the scatter plot of M($\pi^+ \pi^-$) versus
M($p \pi$),
b) the M($p \pi$) mass distribution and
c) the \lamlam mass spectrum.
}
\label{fi:3}
\end{figure}
\begin{figure}[h]
\caption{
a), c), e) and g)
the azimuthal angle ($\phi$) between the two outgoing protons
for the \ppbar, \ppbarpi, \ppbarpipi and \lamlam
channel respectively.
b), d), f) and h)
the four momentum transfer squared ($|t|$) from one of the proton vertices
for the \ppbar, \ppbarpi, \ppbarpipi and \lamlam
channel respectively.
}
\label{fi:4}
\end{figure}
\newpage
\begin{center}
\epsfig{figure=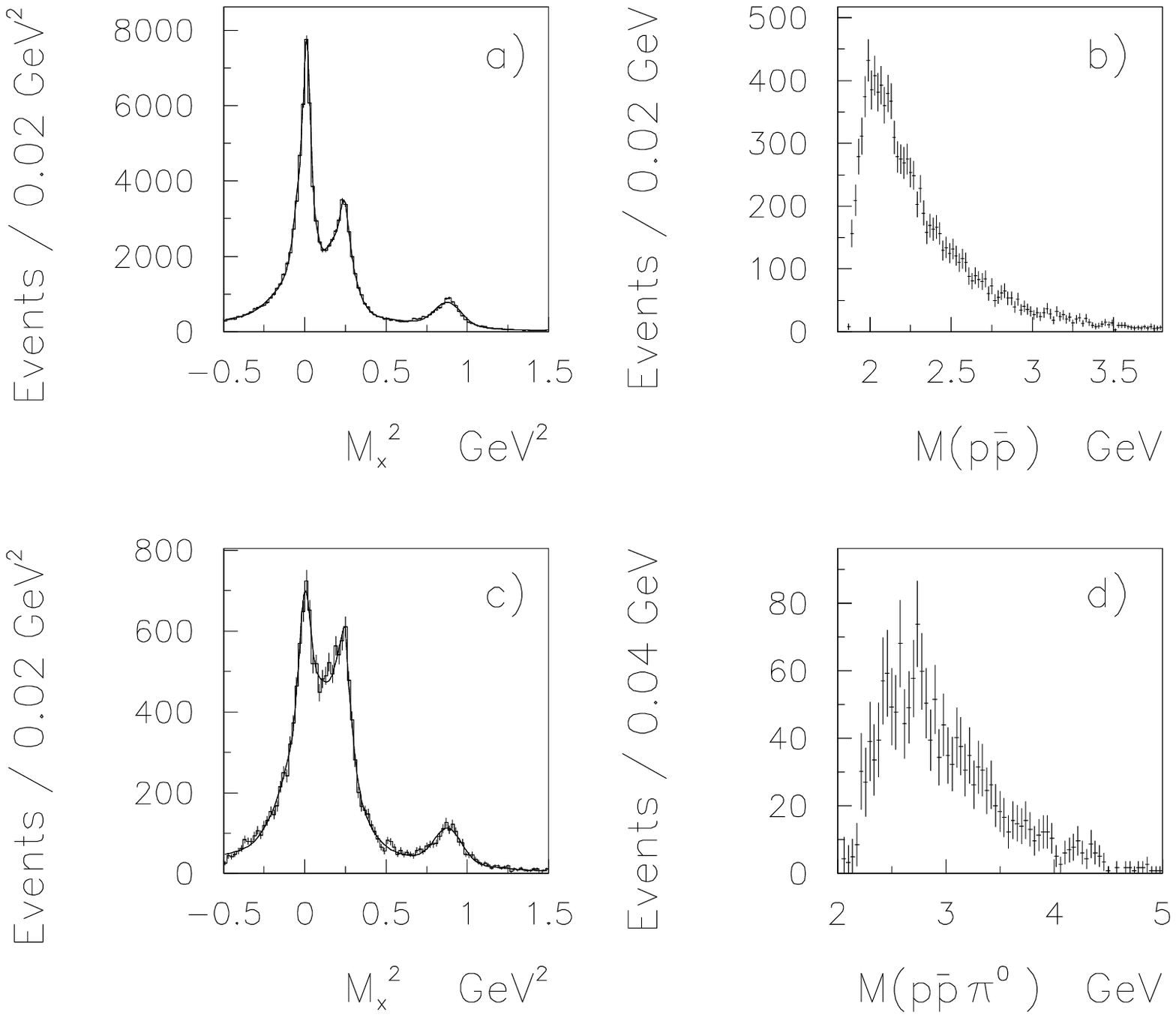,height=22cm,width=17cm}
\end{center}
\begin{center} {Figure 1} \end{center}
\newpage
\begin{center}
\epsfig{figure=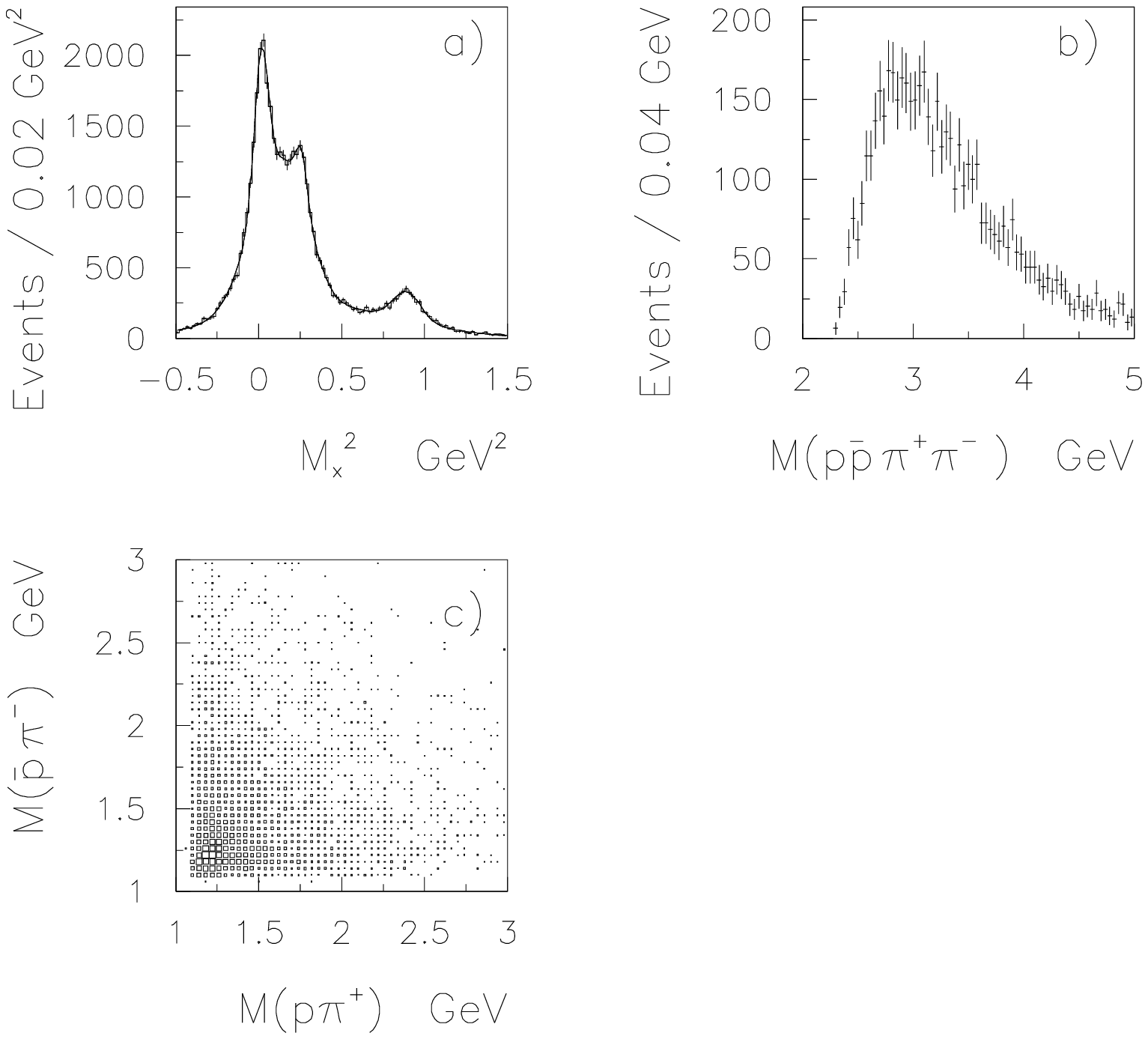,height=22cm,width=17cm}
\end{center}
\begin{center} {Figure 2} \end{center}
\newpage
\begin{center}
\epsfig{figure=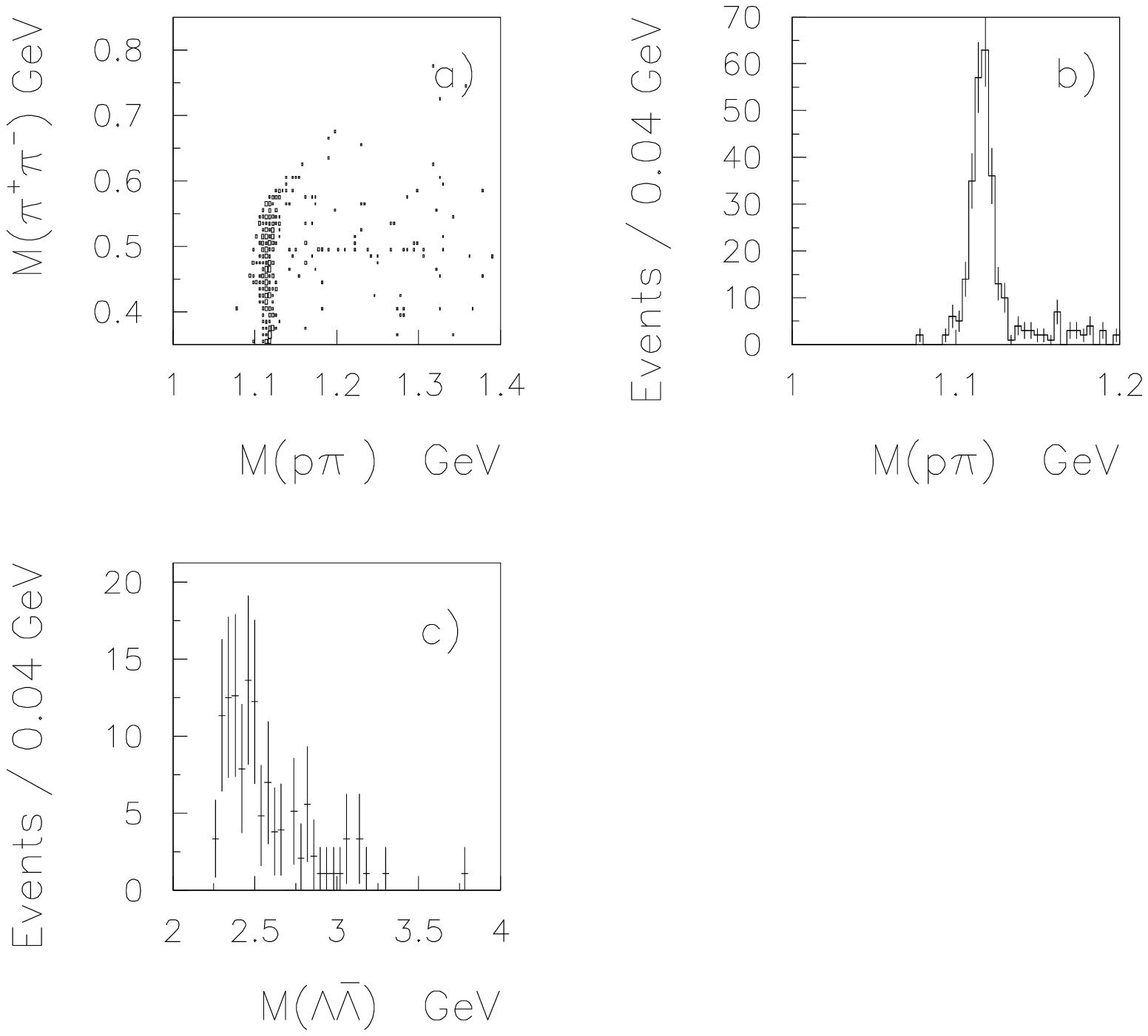,height=22cm,width=17cm}
\end{center}
\begin{center} {Figure 3} \end{center}
\newpage
\begin{center}
\epsfig{figure=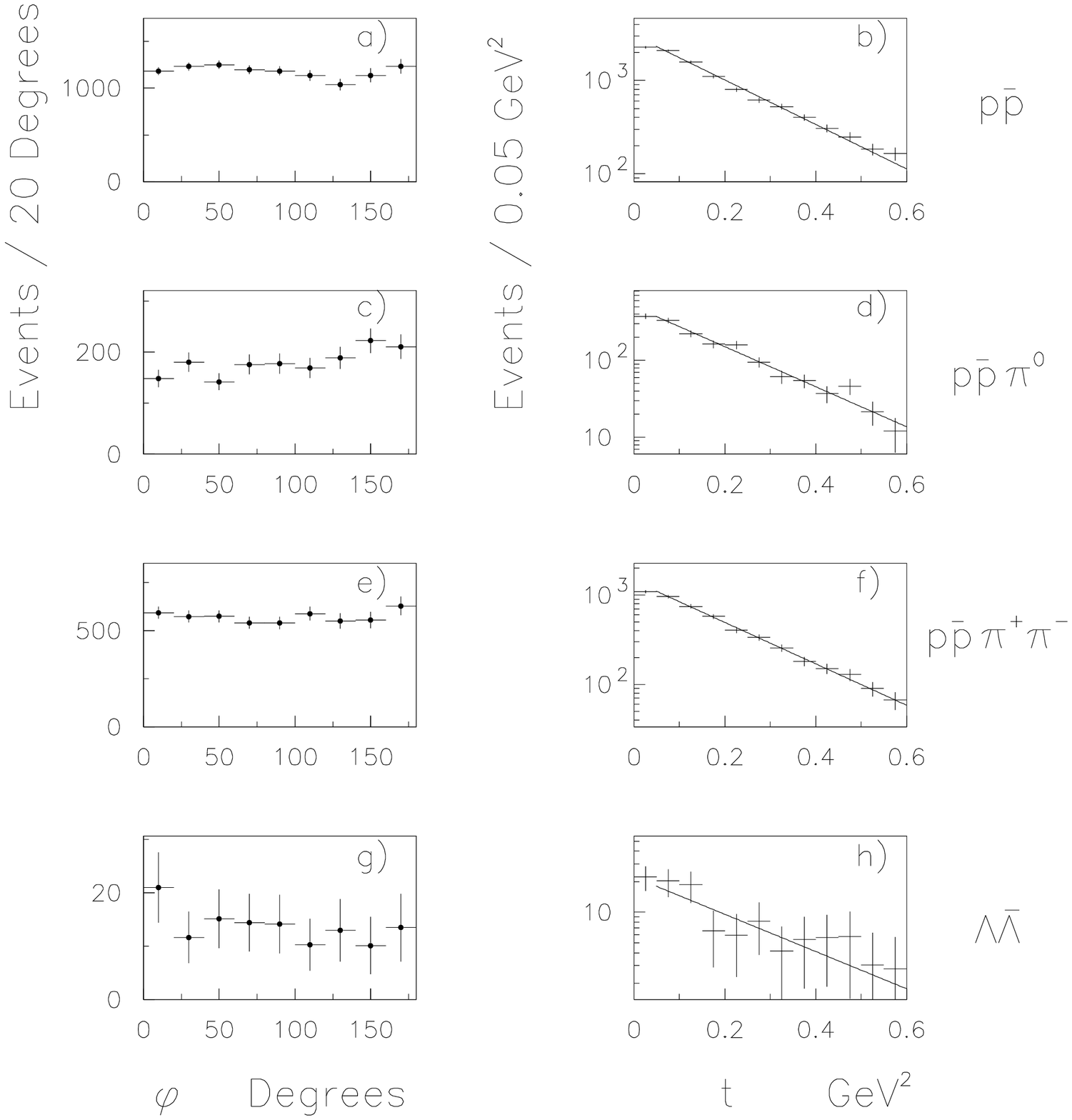,height=22cm,width=17cm}
\end{center}
\begin{center} {Figure 4} \end{center}
\end{document}